# NON-STATIONARY DIVIDEND-PRICE RATIOS

## ABSTRACT

Vassilis Polimenis (*) and Ioannis Neokosmidis (**)

Dividend yields have been widely used in previous research to relate stock market valuations to cash flow fundamentals. However, this approach relies on the assumption that dividend yields are stationary. Due to the failure to reject the hypothesis of a unit root in the classical dividend-price ratio for the US stock market, Polimenis and Neokosmidis (2016) proposed the use of a modified dividend price ratio (mdp) as the deviation between d and p from their long run equilibrium, and showed that mdp provides substantially improved forecasting results over the classical dp ratio. Here, we extend that paper by performing multivariate regressions based on the Campbell-Shiller approximation, by utilizing a dynamic econometric procedure to estimate the modified dp, and by testing the modified ratios against reinvested dividend-yields. By comparing the performance of mdp and dp in the period after 1965, we are not only able to enhance the robustness of the findings, but also to debunk a possible false explanation that the enhanced mdp performance in predicting future returns comes from a capacity to predict the risk-free return component. Depending on whether one uses the recursive or population methodology to measure the performance of mdp, the Out-of-Sample performance gain is between 30% to 50%.

* Aristotle University of Thessaloniki, polimenis@econ.auth.gr

** Aristotle University of Thessaloniki, ineokosm@econ.auth.gr

# I. INTRODUCTION

The ability to forecast financial returns, using well-defined fundamental ratios, has always been regarded as the most significant question for asset allocation, and one of the most important issues in the entire financial economics. Fama and French (1988) have shown how short horizon same-direction forecastability combined with a highly *persistent* forecasting variable mechanically generates substantial return *predictability at a long horizon*. From early on, dividend yields attained special importance as a forecasting variable due to the straightforward participation of the dividend yield in return formation, and its highly persistent dynamics which could provide predictability in long forecasting horizons.

By extending the Campbell-Shiller (1988) approximation to long horizons, one must conclude that long-run return and/or dividend growth predictability should coincide with the variability of the log dividend-price ratio (dp)[1]. This finding is based on two main assumptions, a) the stationarity of dividend yields and b) the assumed ability to recursively extend the CS approximation to infinity. To move beyond this framework, Polimenis and Neokosmidis (PN16) argue that - what empirically emerges in the data as - a slope differential between dividends and prices is probably due to changes in dividend policy. Therefore, to pragmatically approach dp as a forecasting variable we need to move away from the assumption of dividend yield stationarity. Instead we assume a deterministic long run equilibrium relation between dividends and prices of the form $d_t = \alpha + \beta p_t$ and allow the data to reveal the true cointegration vector [1, -β].

In this paper, we extend the Polimenis and Neokosmidis (2016) findings in two main directions. Firstly, we improve the analysis by directly using the approximate relation in Campbell and Shiller (1988), to form weighted long-run returns (wr) in our predicting regressions. The CS approximation is important for our investigation as it directly ties current dividend-price ratio to future returns and growth. Furthermore, the CS-relation is important as it represents a robust way to quantify the basic forward-looking pricing principle: high prices today that are not followed by robust growth will either lead to low future returns or be part of a rational bubble.

Second, we perform a set of important robustness checks as follows

a) by comparing the performance of mdp and dp in the period after 1965 (panel B in our tables) we are not only able to establish the robustness of our early findings, but also to debunk the false line of thought that the enhanced mdp performance in predicting future returns comes from a capacity to predict the (not so interesting) return component from money invested in risk free securities

b) by utilizing a competing econometric procedure the *Autoregressive Distributed Lag (ADL)* method to estimate the trend deviation among dividends and prices (mdp′) and

c) by utilizing monthly dividend reinvestment to calculate annual dividend-yields (d*p).

The rest of the paper is organized as follows. In the next subsection we economically discuss the non-stationarity of dp. In Section II, we present the data and the two methodologies for estimating the trend deviation of a long-run relationship between dividends and prices, and form the modified dividend price ratios (mdp and mdp′). We then move on in Section III to present in-sample predictability for both modified ratios and then use the Campbell-Shiller approximation to perform multivariate analysis where both dp and mdp are present on the right-hand side. In the final section IV we present the out-of-sample performance of mdp and mdp′. Section V concludes.

---

[1] In this paper lowercase letters always denote logs: $d_t = logD_t$, $p_t = logP_t$, and $r_t = logR_t$

## A. NON-STATIONARITY OF THE DIVIDEND YIELD

Econometrically, most researchers argue that dp is a stationary process based on *infinite* sample or *asymptotic* arguments, and take dp stationarity as a given assumption. But in practice, neither the data sets that we use, nor the time horizons that we use to evaluate our models' performance are infinite. At the same time, most empirical studies on return predictability, cannot reject statistically (if not economically) the hypothesis of the presence of a unit root in the dividend-price ratio (Goyal and Welch, 2003; Lettau and Van Nieuwerburgh, 2008; Lettau and Ludvigson, 2005 among others).

*** insert Table 1a around here***

We can see from summary statistics presented in Table 1a, that the dividend-price ratio dp has an autocorrelation φ=.87 (or .93 when monthly dividends are reinvested). Clearly, unit root tests have not enough power for such high φ values. Furthermore, it is known, that typical estimation methods will tend to highly underestimate true persistence in finite samples. In the following sections, we present robust econometric evidence against the stationarity of the classical dp. Not only is stationarity rejected via a straightforward ADF testing for dp, but using the more powerful test of a restriction on the cointegration vector for d and p we reject the hypothesis that log-dividends and log-prices are linked with a long run relationship of the form (d-p).[2]

The economic requirement that stock prices cannot be far from corporate fundamentals for prolonged periods has often been interpreted in a strict sense that the log dividend-price ratio is stationary either in the full sample or at least for large subperiods. The classical line of reasoning about dividend yields is that dividends should represent a more or less "fixed" fraction of earnings, and earnings should represent a more or less "fixed" fraction of prices. This has led most contemporary literature to de facto assume classical dp ratios are *stationary processes* and do not include trends. This is an economic requirement, that depends on the particular sample used, rather than a hard fact. Finance officers have large discretion over payout options, and might impart unexpected structure into the dynamics of the dividend yield.

The fact that, over any finite period of time, dividends (and dividend growth) can be arbitrary, and delinked from asset prices, means we should neither be dogmatic about the time series properties of the dividend yield nor about its inability to predict dividend growth. Yet, generally speaking, both academia and practice have avoided tackling head-on the possibility of non-stationary dynamics in valuation ratios such as the dividend-price ratio, despite the fact that the hypothesis of a unit root in long horizon samples cannot be statistically rejected. The economical source of such non-stationarity in dividend yields is not easily understood. It could be the result of changes in dividend policy such as *dividend smoothing*, use of *share repurchases* in lieu of cash payments, or it could be induced by other changes of investors' attitudes toward dividends and taxes.

In any case, such changes in dividend policy will emerge in the data as a *slope differential* between dividends and prices. PN16 propose that we move away from dividend yield stationarity, by assuming a deterministic long run equilibrium relation between dividends and prices as the next logical step still satisfying a "fundamentals" based asset pricing philosophy. They then modify the dividend-price ratio by relaxing the stationarity assumption for the classical $dp_t$, assume a deterministic long run relation between dividends and prices (i.e. a cointegration vector of the form $d_t = \alpha + \beta p_t$) and allow the data to reveal the true cointegration vector $[1\ -\beta]$.

---

[2] That the [1 -1] vector spans the cointegration space.

Having formed the above long-run relation, PN16 define the modified dividend-price ratio as the stationary cointegration error of this long-run equilibrium, $mdp = d - \beta \cdot p$. In PN16 we argue that β is the unique population parameter that fine tunes the dividend-price relationship by revealing a small non stationary I(1) trend deviation between dividends and prices. Since the classical dp can be thought as the modified ratio, mdp, plus a noise trend, the modified ratio (mdp) is more informative than its non-stationary counterpart

$$dp_t = mdp_t + (\beta - 1) \cdot p_t$$

Since the modified ratio does not de facto assume an econometrically unreliable rejection of the non-stationarity null for dp, it presents a more reliable alternative, which allows for a richer representation of the d.g.p. Also, at φ=.70, mdp still has enough persistence in order to provide forecastability in long horizons.

The coefficient β provides the drift ratio between d and p. Roughly speaking, a β<1 implies that dividends have been growing more slowly than prices. Having motivated the possibility for such a slope differential, and thus a non-stationary dp, the important question with respect to understanding the true dynamics of dp is whether such non-stationarity is only due to a deterministic time trend or it includes a unit root. The problem is that, as is now well understood, this question is inherently unanswerable for any finite sample (see Blough 1992) since for any unit root process, and sample size T, there exists a stationary process that is indistinguishable. Another way to understand this issue is that the question of the inclusion of a unit root in the process is equivalent to finding whether the population spectrum at zero is zero or attains any positive value. This is clearly unanswerable, since in any sample there is no information about cycles of a period larger than the sample size. A realistic target for the financial economist should rather be to describe the data in a parsimonious way with low order autoregressions, since they are easier to estimate than high order moving average processes.

PN16 show that an investor who employs the modified ratio (mdp) will improve his Out-of-Sample forecasting of 3-, 5- and 7-year returns with an $R^2_{OS}$ of 7%, 26% and 31% respectively. Furthermore, an investor who has seen enough of the small (due to super-consistency) required early sub-sample to reliably infer population values for the cointegration coefficient between d and p, will improve his 5- and 7-year returns forecasts by an astonishing $R^2_{OS}$ of 49%, and even attain a 3-year $R^2_{OS}$ of 34%.

It is well known that existing breaks will lower the power of unit root tests (Perron (1989)), thus making stationary processes with breaks difficult to distinguish from those including a unit root. Thus a competing approach that may also produce a good in-sample fit to the data is to allow for occasional breaks to the levels or the slope of an otherwise stationary process (e.g Fama and French (2002) consider a mean reverting dp within different regimes), Yet, allowing for breaks that are impossible to predict ex ante has little value for return predictability and forecasting, and will produce a weak out of sample $R^2$ (see Lettau and Van Nieuwerburgh (2008)). This is clearly not the case with our parsimonious approach that produces significant out-of-sample forecasting gains.

# II. CONSTRUCTION OF THE RATIOS

High quality return data for the S&P 500 index, with and without dividends, are available from CRSP since 1926. As in PN16, our full sample[3] spans the most recent 87-year period that ranges from January of 1926 to December of 2012. We extend PN16 by presenting a new analysis based on the 1965-2012 subperiod. We only use nominal data throughout the paper. [4]

## A. CONSTRUCTION OF THE CLASSICAL DIVIDEND-PRICE RATIO

Below we show how we use (total and ex-dividend) monthly returns' data for the S&P 500 to formulate annual dividend and price level series, and the classical dividend-price ratios (dp and d*p).

If we denote by $D(t)$ the monthly dividend for month t, by $R(t)$ monthly gross return, $R(t) = \frac{P_t + D(t)}{P_{t-1}}$, and the monthly return due to price gain alone (i.e. without dividends) $X(t) = P_t/P_{t-1}$ respectively, the monthly dividend for month (t) is given by[5]

$$D(t) = \left(\frac{R(t)}{X(t)} - 1\right) P_t$$

To cancel dividend seasonality, we employ an annual horizon when constructing the log dividend-price ratio. Depending on how one forms annual dividends at the end of month t, from the 12 preceding monthly dividends, the dividend price ratio may be computed with two different methodologies. If we denote with $D_t$ the ending at month t annual dividend, the most typical annualized dividend computation is via $D_t = \sum_{i=0}^{11} D(t-i)$. This leads to the most commonly used dividend-price ratio based on the following computation

$$DP_t = \frac{D_t}{P_t} = \frac{\sum_{i=0}^{11} D(t-i)}{P_t}$$

We finally form dp as the difference of logs

$$dp_t = d_t - p_t = \log(D_t/P_t)$$

A less common method is to form a dividend-price ratio by immediately re-investing interim dividends as they become available. This technique is more appropriate from a conceptual point of view, but also transfers to dividends some of the market volatility for the year.[6] We use d* for annual reinvested dividends and d*p for the reinvested dividend yield via d*p=d*-p.

---

[3] The data are from the Goyal and Welch database, available at http://www.hec.unil.ch/agoyal
[4] See also the discussion in Campbell and Shiller (1988) and Cochrane (2008). Engsted and Pedersen (2010) find that long-horizon predictability depends on whether returns and dividend growth are measured in nominal or real terms.
[5] We should not confuse monthly and annual series. We will always use R(t) for the monthly return, with $R_t$ the annual return formed at the end of month t.
[6] Chen (2009) finds that the annual (from monthly) dividend construction can have significant implications on estimated dividend growth predictability, as the reinvestment assumption makes dividends inherit a lot of the intra-year realized return volatility.

# B.    CONSTRUCTION OF THE MODIFIED DIVIDEND-PRICE RATIO

In this section and the next the two econometric techniques we use in forming the modified dividend price ratios ($mdp_t$ and $mdp_t'$ respectively) are described.

Firstly, as in PN16, we estimate the modified dp ratio using the multivariate Johansen (1991) methodology. At its core, the Johansen method uses the size of the eigenvalues of an impact matrix C = AB′ to infer its rank. Specifically, the method infers the cointegration rank by testing the number of eigenvalues that are statistically different from 0. Although the method may appear to be very different from the Engle-Granger (1987) approach, Johansen's maximum likelihood approach is essentially a generalization of the augmented Dickey-Fuller (1979) test for unit roots in many dimensions.

If **w** is a two-dimensional vector $\mathbf{w_t} = [d_t\ p_t]'$, and there exists a cointegrating vector **b**, then $\mathbf{b'w_{t-1}}$ is the "error" in the data that quantifies a deviation from the stationary mean at time t−1. Error correction in our context manifests itself as the tendency of a cointegrated dividend and price series to revert to a common stochastic trend. The *modified* dividend-price ratio ($mdp_t$) is defined as the trend deviation from the established long-run equilibrium between dividends and prices

$$\mathrm{mdp}_t = d_t - \beta p_t \qquad (1)$$

The dividend and price series correct from the "disequilibrium" that mdp represents at rates captured by a vector of their specific *adjustment speeds* **a**, thus forming a multiplicative error-correction term $\mathbf{ab'w_{t-1}}$ that needs to be added to a simple VAR model explaining jointly price change (Δp) and growth (Δd) dynamics and thus produce the so-called vector error-correction VEC(q) model

$$\Delta \mathbf{w(t)} = \sum_{i=1}^{q} B_i \Delta \mathbf{w(t-i)} + \mathbf{a}(\mathbf{b'w_{t-1}} + c_0) + \mathbf{c_1} + \mathbf{u(t)} \qquad (2)$$

The Johansen test for *deterministic* cointegration above[7] addresses many of the limitations of the Engle-Granger method. As a two-dimensional vector **w** = [d p]′ is used here, the main benefit of the Johansen method is that it avoids the two-step procedure,[8] and thus provides a framework for testing restrictions on the cointegrating relations vector **b** (and the adjustment speeds **a**) in the VEC model.

While it is true that the trace and maximum eigenvalue co-integrating rank tests in Johansen are derived under the assumption of Gaussian iid innovations, it has been shown that the standard rank tests based on asymptotic critical values remain asymptotically valid even in the presence of conditionally heteroskedastic shocks,[9] and the trace statistic is more robust to both skewness and excess kurtosis. The trace tests[10] show that the series are cointegrated with a cointegration relationship of the form,

---

[7] Equation (2) may be slightly different depending on whether we assume that are no intercepts or trends in the cointegrating relations and there are no trends in the data. Here we assume that the log series have linear trends but the cointegration relationship contains only a constant. This specification is a model of deterministic cointegration, where the relations eliminate both stochastic and deterministic trends in the data.
[8] A concern related to the Engle-Granger method is that it is a two-step procedure, with the 1st regression employed to estimate the residual series, and the 2nd regression to test for a unit root. Errors in the 1st estimation are automatically carried into the 2nd. Moreover, the estimated, and not observed, residuals require different tables of critical values for standard unit root tests.
[9] Seo (1999), Boswijk (2001), Kim and Schmidt (1993) among others suggest that the standard procedures are asymptotically valid both for unit root and cointegration. Nevertheless, the unit root tests have size distortions in small samples. Rahbek, Hansen and Dennis (2002) find that the usual procedures in order to test for cointegration based on the multivariate settings are asymptotically valid in the presence of multivariate conditional heteroscedasticity (for further analysis on this concept see Harris and Sollis (2003)).
[10] Maximum-Eigenvalue test statistics have also been calculated and support similar findings.

$$\text{mdp}_t = d_t - 0.8017 p_t \qquad (3)$$

## C. THE AUTOREGRESSIVE DISTRIBUTED LAG (ADL) METHOD

In the general setup, one may have n-dimensional time series and there may be multiple cointegrating relations among the variables. In order to provide comprehensive testing in the presence of multiple cointegrating relations,[11] the Johansen test estimates a Vector Error Correction (VEC) model. When a simple dynamic model such $d_t = b_0 p_t + b_1 p_{t-1} + a_1 d_{t-1} + e_t$ is a sufficient representation of the underlying relationship between d and p, super-consistency ensures it is asymptotically valid to omit the short run elements of the dynamic model. If a complicated dynamic relation between d and p exists, ignoring the full dynamic model will push more dynamic terms into the residual e, so that it exhibits significant autocorrelation.

A different procedure, the *Autoregressive Distributed Lag (ADL)* method, is to use a more generous dynamic model to estimate the long run relation between dividends and prices

$$A(L)d_t = a_0 + B(L)p_t + u_t$$

with A(L) and B(L) the polynomial lag operators $A(L) = 1 - a_1 L - a_2 L^2 - \cdots a_p L^p$ and $B(L) = b_0 + b_1 L + b_2 L^2 + \cdots b_q L^q$. To estimate the trend deviation between annually summed dividends (d) and prices (p), we employ the following ADL model,

$$d_t = a_0 + \sum_{i=1}^{p} a_i d_{t-i} + \sum_{j=0}^{q} b_j p_{t-j} + \varepsilon_t \qquad (4)$$

Which is then transformed [see the appendix] to a long run estimated solution of the type

$$d_t = \hat{\alpha} + \hat{\beta} \cdot p_t$$

The great strength of using the ADL approach is that it does not need to assume anything about the form of the dynamic adjustment process between d and p. Using the β estimated by the ADL methodology (3) in (1) allows us to formulate an alternative modified ratio,

$$\text{mdp}'_t = d_t - 0.8548 p_t \qquad (5)$$

## III. IN-SAMPLE PREDICTABILITY

In this section, we present the main univariate and multivariate forecasting regressions based on the classical dividend-price ratios (dp and d*p) and the two modified ratios (mdp and mdp′) respectively. We formulate continuously compounded returns, equity premia, and dividend growth for 1, 3, 5 and 7-year horizons (h = 1,3,5,7) using monthly S&P 500 data. We consider the full sample, spanning the period (1926-2012), in Panel A. Additionally, we use an economically meaningful subsample period running from 1965 to 2012 to have a clearer picture of the dynamics in a more recent environment. Long horizon series are formulated using a rolling window of overlapping monthly observations. Standard errors are GMM corrected based on the Hansen-Hodrick formula.[12]

---

[11] This is of no consequence here as we are dealing with a 2-dimensional vector

Panel A of table (2a) presents the full sample univariate results for all ratios. Long-horizon forecasts are the mechanical result of short horizon same-direction forecastability combined with a highly persistent forecasting variable. The persistence of a predictor variable leads to increased slope coefficient for longer horizons. This is a well understood effect in the literature starting as early as Fama and French (1988). As a manifestation of these predictability mechanics, in Table 2a we have increasing slope coefficients and $R^2$ with horizon as the mechanical result of the highly persistent forecasting variables. All ratios can predict returns and equity premia for all horizons, but the modified ratios can achieve better results for slope, t-stats and $R^2$ values respectively.

***insert Table 2a around here***

The new insight in PN16 is that part of the high persistence of a non-stationary dp is due to the small embedded unit root in $dp_t = mdp_t + (\beta - 1)p_t$. The extra dp persistence though, unlike the "useful" persistence of the stationary mdp, carries no real predicting power. In PN16 we argue that the true forecasting horizon is determined by the lower mdp persistence. The artificially longer useful predicting horizon for dp, that one gets by mechanically extending short period dp predictability into the distant future, is an artifact of the non-stationary noise embedded in dp and of no real forecasting value.

For all return horizons, both modified ratios (mdp and mdp′) achieve impressive improvements over both classical ratios in all three dimensions: slope size, significance (t-stats), and long-return explanatory power ($R^2$). Not only modified ratio performance strictly dominates classical ratios in all horizons, but furthermore, this modified ratio dominance gets more pronounced with an increasing horizon.

To better understand how mdp works, PN16 plot both ratios against future realized returns. Figure (1) plots the 5-year future realized long run returns against current dp and mdp levels. Note the surprising ability of mdp to avoid the excessively low dp print in the early 2000s. This happens because, in a world where some dividend policy trend (e.g. an increasing use of share repurchases) has induced non-stationarity in dp yields, mdp captures the true deviation from long run equilibrium between prices and fundamentals, by mechanically factoring out the non-stationarity inducing dynamic in the dividend yield.

***insert Figure 1 around here***

## A. EVIDENCE FROM THE MORE RECENT HISTORY

The strong over-performance of mdp over dp, in predicting future returns, is considerably toned down when using mdp in explaining equity premia. While still strong, in the full sample, the performance of mdp in forecasting equity premia is comparable to the performance of the classical ratios. Since total equity return is composed of the risk-free return plus the realized equity premium, a false line of thought with mdp could be that the enhanced performance in predicting future returns comes from a capacity to predict the (not very interesting) return component from money invested in risk free securities. Indeed, as shown in Panel A of Table 2b, in the full sample, for all tested horizons, risk free returns are forecasted by mdp but not dp.

---

[12] In order to correct heteroskedasticity and correlation effects Newey-West estimates of the standard errors have also been tried with no change on the significance of the findings.

***insert Table 2b around here***

As shown here, this line of criticism is not valid. To debunk this issue, and better understand the sources of the enhanced mdp performance, we perform the analysis for the more recent sample 1965-2012. The first observation is that while in the truncated sample (panel B) classical dp $R^2$ for explaining returns are significantly higher, the modified ratios retain a clear superiority. For example, while 5- and 7-year classical $R^2$ increase from 19% and 25% to 37% and 47% respectively, in the same, post-1965, period the mdp return $R^2$ becomes 54% and 72% respectively.

The second finding in Panel B is that while both the classical and modified ratios can predict equity premia for all horizons, mdp clearly dominates with an $R^2$ as high as 44% for the 7 years ahead premium. Classical ratios can only attain an $R^2$ of 17% at the same horizon. Furthermore, while for the full sample classical ratios cannot forecast risk free yields, in the recent subsample the classical ratio $d^*p$ not only forecasts risk free rates but does better than mdp. While for a 3-year horizon, dp explains 60% of future risk-free returns, mdp only captures 35% of this variability. In the 5 and 7-year horizons, classical dp captures more than 70% of future risk-free returns, while mdp captures only around 50%. So, the forecasting gain of mdp comes despite classical ratios better explaining future risk-free returns.

It is important to economically discuss the positive correlation of both (classical and modified) ratios with future risk-free returns post 1965. We know that, given the high persistency of short term yields, T-bill returns are highly forecastable. If interest rates (and hence one-year risk free returns) are currently low, they are likely to remain low for the next years as well. If companies that consistently pay dividends attract a certain type of investor (clientele) then such companies can get away with low dividend yields when such low payouts coincide with low current and future (due to their high persistency) risk free yields. In such low-yield states of the economy, income seeking investors will not allocate their portfolios out of low dividend yield stocks because they have nowhere to go.

As the deviation between d and p from their common trend, mdp factors out a tendency to pay lower dividends over time (probably also due to the use of share repurchases) and thus consistently captures true dividend-yields at any point in time. By not properly factoring out the continuous move out of dividend payments (and into share repurchases) after the 80s, classical dp goes very low in the 90s and around 2000 but fails to predict a catastrophic price correction.

The superiority of mdp over dp in capturing future return variability is also shown when we run multivariate regressions at 5 and 7-year horizons with both dp and mdp present on the right-hand side (see Table 2c). When both dp and mdp compete to explain future return realizations, for all horizons mdp comes out significant while dp becomes insignificant (only marginally significant for the 7-year horizon in the limited post-1965 sample).

***insert Table 2c around here***

## B. THE CS APPROXIMATION AND THE NON-STATIONARITY OF THE DIVIDEND YIELD

In this section we use a multivariate analysis to investigate the sources of finite horizon dp variability in the presence of mdp. This analysis can be delivered using the well-known Campbell-Shiller (CS) relationship[13]. The CS relation is important as it analytically describes and quantifies a fundamental pricing principle: in an efficient market, high stock prices that are not due to a strong growth expectation, are either due to low discount rates, or are part of a rational market bubble

$$pd_t \approx E\sum_{j=1}^{h} \rho^{j-1}\Delta d_{t+j} - E\sum_{j=1}^{h} \rho^{j-1}r_{t+j} + \rho^h E pd_{t+h} \quad (6)$$

The powerful feature of the CS-relation is that it holds not only ex ante, as an expectation about the future, but is also valid ex-post on a *path-by-path* basis.

$$dp_t \approx \sum_{j=1}^{h} \rho^{j-1}r_{t+j} - \sum_{j=1}^{h} \rho^{j-1}\Delta d_{t+j} + \rho^h dp_{t+h} \quad (7)$$

The current level of the dividend-price ratio completely determines this linear combination between *realized* long run returns, growth and future dividend yields; i.e. by attending dp today we know completely the investors' expectations for this sum. As the relation holds ex post, no new information can disturb this sum, and inclusion of any extra information $I_t$ about long run terms can only re-arrange these terms in a way that respects the CS-sum as it is calculated by the current dp.

For example, if new information in another variable leads to higher long run discount rates, it will either have to also forecast an increased long run dividend growth and/or higher future prices. Cochrane (2011) suggests taking regressions of long run terms on the dividend-price ratio in order to reveal the source of dp variation. As the horizon increases, the source of dp variation needs to correspond to fundamentals (i.e. returns and/or dividend growth).

Unlike the auto term $\rho^h dp_{t+h}$ that vanishes by being multiplied by $\rho^h$, when expanding the CS relation in a long horizon, the approximation error does not vanish, but rather successive errors get cumulated[14] at decreasing weights. Thus, a highly persistent dp that deviates far from the expansion point in a particular sample may produce significant cumulative approximation residuals in the long horizon application of the formula. The theoretical properties of the error induced by the long horizon application of the approximation remain largely unknown. The limited power in detecting near non-stationarity in dp, and the uncontrollable error that a unit root in dp might inject into the approximation, is a strong argument that favors the use of limited (5- or 7-year) instead of infinite horizons when analyzing the performance of the CS-approximation.

***insert Table 3 around here***

Table (3) shows the CS analysis of the total dp variation for 5 and 7 years ahead based on *weighted* forecasting regressions of future returns, dividend growth and the auto-correlated term on dp. We use annual data that are constructed from the original monthly observations. One may in principle use the CS approximation in monthly horizons, but would then have to deal with the problem of strong seasonality in dividend payments. As we can see, discount rate variability can roughly account for 37% and 46% of the total dp variation for 5 and 7 years ahead respectively. At these horizons the auto-correlated component still has a strong effect on the dp roughly explaining another 60% and 47% of its variability respectively.

---

[13] In this section, to enhance readability, we denote the dividend-price ratio as ($dp_t$), while we always run regressions based on ($d^*p_t$) constructed for an annual horizon. This is because, in order to retain the CS identity at an annual horizon, it's necessary to use the reinvested dividends ($d^*$) for the dividend-price ratio and dividend growth construction.
[14] This is already discussed in the original Campbell and Shiller (1989) paper. In order to evaluate the magnitude of the approximation error, Campbell and Shiller compare actual returns to the ones predicted by their approximation and find it small and almost constant. Yet, their empirical analysis is using data up to 1986, and most of the interesting behavior of the dp ratio occurs around 2000.

An important observation is here in order: If the CS identity was holding exactly, the three slope coefficients should sum up to one. The departure from the theoretical 100% limit is due to the approximation error in the CS sum, and the size of this deviation is a way to quantify the approximation error in the particular horizon for that path. For example, at a 7-year horizon, the slope sum is around 96%, implying a significant approximation error. Actually, the fact that the slope coefficients do not sum up to one is even more significant, because it is evidence that the CS error is not only sizeable but also exhibits significant variation and correlation with dp itself. Measuring the slope sum deviation from 1 is actually the proper way to measure the economic significance of the CS error. On the contrary, when measuring the modified slopes, in Table (4), the respective sum is very close to zero and always insignificant. This is evidence that mdp, unlike dp, is uncorrelated to the CS error.

***insert Table 4 around here***

In Table (4) we present the findings of a multivariate analysis with the modified ratios mdp (and mdp′) placed along the classical dividend price ratio ($d^*p$). We see that for both horizons the modified dividend-price ratio completely drives out the classical dividend-price ratio. While in all cases the modified ratio attains a return coefficient ($c_r$) close to one and very significant, the dividend-price ratio slope ($b_r$) is always insignificant and close to zero. This is a significant conceptual departure from the prevalent point of view that most dp variability captures variation in discount rates, and therefore needs to be carefully analyzed.

More specifically, the contemporary doctrine is that when forming (sufficiently) long-horizon weighted predictive regressions all dp variability is explained by changes in long run returns; that is since $(\rho\phi)^h \to 0$, $b_r \to 1$, $b_g \to 0$.[15] In a stationary dividend-price ratio world, this erroneous hypothesis taken to a limit of a very long horizon would imply that since by construction[16] $(c_r - c_g + c_u = 0)$, when failing to predict dividend growth $(c_g = 0)$, all an extra forecasting variable can do is to capture the term structure of long run discount rates. Table (4) clearly cannot be explained along these lines, and the analysis must be differentiated as follows. The picture in Table (3) is that for medium term horizons, some of the dividend-price variation is driven by changes in long run discount rates and the rest is driven by long-run autocorrelation (the auto term). When the modified ratio is added in the picture, in a multivariate setting, the superiority of its filtered information is so powerful that the dividend price ratio completely fails to bring any extra information about future returns to the table ($b_r = 0$). Thus, in the presence of the modified ratio the only thing left for the classical dividend-price ratio to predict is its forward looking value through its autocorrelation; i.e $b_r = b_g = 0, b_u = 1$.

The problem with the third slope coefficient ($b_u$) is that, in a classical environment (i.e. stationary dividend-price ratio), this term will have to go to zero for long horizons (autocorrelation gets multiplied by $\rho^h$). But in a nonstationary dividend-price world as the one we assume (and we have econometrically failed to reject), the picture is more complicated since the error in the CS approximation is not bounded anymore. Since mdp is stationary by construction, if $(b_u)$ is to survive for long horizons it can only do so through the nonstationary "noise" in dp (i.e. $(\beta - 1)p_t$.) But if $b_u = 1$ for long horizons, and since the only thing that survives in large horizons is the nonstationary I(1) component in dp, we can conclude that the right economic picture for the long horizon weighted forecasting slope coefficients is probably better captured by $(b_r = b_g = 0, b_u = 1)$ and $(c_r = 1, c_g = 0, c_u = -1)$. Then, the modified ratio captures economically all the fundamental variation of dp and leaves the "noise" part to explain the variation of dp which comes from the auto "bubble" component.

---

[15] Here we use g to denote annual dividend growth Δd.
[16] With $c_r, c_g, c_u$, the mdp slope coefficients for returns, dividend growth and the auto factor ($\rho^h dp_{t+h}$) respectively.

A critical issue is to understand where the enhanced ability of the modified ratios to predict long run returns comes from. One method to shed some light is to break down future returns on the risk free returns plus the equity premia ($r_t = re_t + rf_t$). Long run returns from investing in T-bills should be much easier to predict that future equity premia, since it is well known that T-bill returns are predictable due to the slowly moving nature of yields. We run multivariate long run forecasting regressions based on the expanded CS approximation,

$$dp_t \approx \sum_{j=1}^{h} \rho^{j-1}(re_{t+j} + rf_{t+j}) - \sum_{j=1}^{h} \rho^{j-1}\Delta d_{t+j} + \rho^h dp_{t+h} \qquad (8)$$

and present the findings in Table (4b).

For a 5-year horizon, the .87 (1.02) forecasting slope breaks down to a .51 (.62) slope in capturing future equity premia, plus a .37 (.40) slope explaining T-bill returns for the modified ratio $mdp_t$ (respectively $mdp'_t$). Not surprisingly, as the horizon gets longer to 7 years ahead, an even larger fraction, of the forecasting multivariate slope .93 (1.05) of $mdp_t$ ($mdp'_t$), equal to .43 (.50) is due to its power to forecast future T-bill returns. At 7-years ahead, the remaining .50 (.55) slope on the modified ratios captures future equity premia.

***insert Table 4b around here***

## IV. OUT-OF-SAMPLE PERFORMANCE

In this section, we extend the PN16 Out-of-Sample (OS) analysis for all classical and modified ratios and both periods. As usual, the evaluation is done by comparing against the forecasting ability of a simple benchmark for a real-time investor. Campbell and Thompson (2008), who summarize the forecasting power for a pool of common financial and accounting variables, propose the use of Out-of-Sample coefficient of determination which is computed via the statistic

$$R^2_{os} = 1 - [\sum_{k=1}^{\tau}(r_{t+k} - \hat{r}_{t+k})^2 / \sum_{k=1}^{\tau}(r_{t+k} - \bar{r}_{t+k})^2]$$

This measure compares the OS performance of a predictor variable that predicts a return $\hat{r}$ against the "simplistic forecast" benchmark that utilizes the simple average of past returns $\bar{r}$ as forecast. The OS coefficient of determination $R^2_{OS}$ effectively asks if we could do a better forecasting job than someone who just expects that returns will always be the same. When compared with the squared Sharpe ratio, a positive $R^2_{OS}$ is directly related to the welfare benefits for a mean-variance investor achieved by using the predictor variables.

We present the Campbell-Thomson OS coefficient of determination, for predicting returns and equity premia for 3-, 5- and 7-year horizons. We divide the data sample in two periods. Initially, we utilize a 15-year minimal slope *estimation* period (1926-1941) or (1965-1980) for the subsample. The remaining sample, extending beyond the estimation period (until 2012), constitutes the *evaluation* period. We choose 15 years for the initial estimation period as it is necessary to have enough initial data to provide reliable slope OLS estimators, and at the same time a large evaluation period for reliable OS appraisal (see the discussion in Welch and Goyal (2008)).

While calculating dp from current data is straightforward, for the econometrician to construct mdp, the true long-run coefficient (β) between d and p needs to be estimated first. On a first approach, the straightforward method is to re-estimate the cointegration coefficient on a *recursive* (R) basis, each time using only data up to a certain point t. This means a two-step procedure for every time t: a) a cointegration coefficient $β_t$ with d and p data only up to time t is estimated, and b) $mdp(R) = d - β_t p$ for the 1…t period is formed, so that finally a forecasting regression of returns against that mdp in 1…t can be run.

***insert Table 5 around here***

As we can see, the classical dp ratio cannot provide positive $R^2_{OS}$ values, meaning that it fails to outperform the simplistic forecast benchmark for all short- to medium-term horizons. To get the dp ratio to (even marginally) outperform the simplistic forecast we need to utilize a long 7-year horizon. On the other hand, the recursively estimated modified ratio provides forecasting benefits as fast as in predicting the 3-year forward return. An investor who employs the modified ratio mdp(R) will improve his Out-of-Sample forecasting of 3-, 5- and 7-year returns with an $R^2_{OS}$ of 7%, 26% and 31% respectively.

Thus, use of the mdp addresses a major weakness in dp, namely its presumed inability in capturing high to medium frequency (i.e. business cycle) variation in expected returns. As the investing horizon gets longer, the modified ratio $R^2_{OS}$ is increasing and reaches a 31% strong gain for seven years ahead. Even a small $R^2_{OS}$ can provide great investment benefits for investors who would erroneously assume that "…returns will be as they always have…" (see Campbell and Thompson 2008; Rapach et.al, 2010).

### A.    PERFORMANCE OF THE POPULATION MDP

Actually, even the strong performance of mdp(R) depicted in Table 5 is conservative, and the true forecasting benefit of mdp for a large sample is probably even higher. Although it is more agreeable from an informational point of view, the recursive procedure carries great sampling errors, and puts the predicting power of mdp(R) at a disadvantage. This happens because, when estimating $β_t$ recursively, we run forecasting regressions of future returns against an mdp proxy, and not against the true mdp that would be produced using the *population* coefficient β.

Furthermore, *due to the super-consistency* of the cointegration estimator only a small early sub-sample is required to reliably infer *population* values for the cointegration coefficient between d and p. Thus, in Table 5 we not only present the recursive performance of mdp(R), but also estimate a *population* (or long-run) mdp(P) where the co-integration coefficient is estimated using the full sample. Effectively, the difference between mdp(P) and mdp(R) measures the forecasting gain for an investor who has seen enough data to recover the population coefficient β. As shown in Table 5, an investor who has seen enough of this early subsample, will actually improve his forecasts for the 5- and 7-year returns by an astonishing $R_{OS}^2$ of 49%, and even attain a surprising 34% Out-of-Sample 3-year $R^2$ statistic.

A concern with evaluating the performance gain of the population mdp(P) is whether a practitioner operating in the early part of our sample, and estimating cointegration coefficients without access to enough historical data, could have exploited the full forecasting power of mdp(P) to his advantage. This "look ahead" concern, when we try to examine the out-of-sample power of our modified ratios, is well documented by Lettau and Ludvigson (2001) in the similar case of evaluating the performance of their cay variable. There is an inherent difficulty in addressing this issue, since subsample analysis (such as out-of-sample forecasting tests) entails a loss of information, and may fail to reveal the full forecasting ability measured with in-sample tests. For reasons explained also in Lettau and Ludvigson (2001), the appropriate estimation strategy for measuring the full forecasting power of the modified ratios, could be to use the full sample, because sufficiently large samples of data are necessary to recover the true cointegration coefficients. Assuming that the investor knows the population coefficient is not a heavy requirement because cointegration coefficients are super-consistent, converging fast to their true values at a rate proportional to the sample size T.

Figure 2 shows a graph of recursively estimated $β_t$ coefficients over the sample. As shown in Figure 2, an investor who has seen as little as 30 years of data may treat estimated coefficients as long-run β values during the second-stage forecasting regressions.

***insert Figure 2 around here***

# V. Conclusion

We have seen that significant forecasting gains are achieved when we employ the modification to the calculation of the dividend price ratio. Besides the econometrics, on a practical level we feel that the true performance gain should be somewhere between the performance of mdp(R) and that of mdp(P); i.e. the two statistics should be viewed more as a low and high limit on the forecasting gain of the dp modification. Depending on whether we use the recursive or population methodology to quantify the performance of mdp, the Out-of-Sample performance gain is between 30% to 50%.

# VI. Appendix

## A. Integration and Co-Integration of the series

In the notation here, $\mathbf{w_t} = [d_t \ p_t]'$ represents the vector of underlying log dividend and price series, and stationarity of the classical dp is robustly tested in a straightforward manner as a restriction $\mathbf{b} = [1 \ -1]'$. To test for cointegration via Johansen, as a first step, the deterministic components which are involved in both the short and long run dynamics and the optimal lag length (q) need to be specified. As shown in (2), the choice here is to consider that the log series have linear trends (captured in $\mathbf{c_1}$) but the cointegration relationship contains only a constant ($c_0$). This specification is known as *deterministic cointegration* between trending series.

For the optimal lag selection, an unrestricted VAR model in levels with a high initial number of auto-regressive lags is first estimated, and then the higher order autoregressive coefficients are tested for significance. Estimating an initial VAR in levels is crucial for the convergence properties of the usual test statistics. An extended reference on this subject can be found in Hamilton (1994, ch.18) and Toda and Yamamoto (1995). Initially, a maximum order of 12 lags is assumed, that is then conditioned down to a more parsimonious representation based on the Hannan-Quinn criterion. The procedure supports an optimal use of 7 lags for the VAR specification, and thus 6 lags in the VECM system.

*** insert Table 1b around here***

The results are presented in Table 1b. The second panel of Table 1b presents results from testing the null restriction that the vector [1 -1] spans the cointegration space based on the Johansen procedure on [d p], which is also strongly rejected. As the Johansen procedure is essentially a multivariate generalization of the augmented Dickey-Fuller test for unit roots, this is more powerful empirical proof of the nonstationary behavior of dp that deals with the low power of unit root tests against highly persistent alternatives[17].

## B. COINTEGRATION VIA THE AUTOREGRESSIVE DISTRIBUTED LAG (ADL) METHOD

The true lag lengths in (4) are not known. If a complicated structure with more dynamic components is used, the possibility of multicollinearity issues arises (a high $R^2$, but imprecise parameter estimation with low t values) even if the model has been correctly specified. If instead a comprehensive representation with fewer dynamic terms is used, it is likely that some residual autocorrelation noise remains. A choice is thus made here to consider three lags for both dividends and prices in the estimation of (4).

The optimal ADL model in (4), with p=q=3, is then transformed to a long run estimated solution of the type

$$d_t = \hat{\alpha} + \hat{\beta} \cdot p_t$$

with $\hat{\alpha} = \hat{a}_0/(1 - \sum_{i=1}^{3} \hat{a}_i)$ and $\hat{\beta} = \sum_{j=0}^{3} \hat{b}_j /(1 - \sum_{i=1}^{3} \hat{a}_i)$.

This transformation of an ADL model to its corresponding long run solution can only be performed under the cointegration assumption between the series. The latter depends on a unit root test of the form $(H_0: (1 - \sum_{i=1}^{3} \hat{a}_i) = 0)$. At 5% significance level, using the critical values of Ericsson and MacKinnon (Ericsson and MacKinnon, 2002) the null is rejected [t-statistic is -3.30]. Therefore, the model can be transformed to its long run solution.

Any hypothesis imposed on the ADL-estimated long run parameters, such as the slope coefficient (β), can be tested asymptotically under the standard distributions. We find that the null hypothesis of [1, -1] is rejected with a t-statistic of (-2.81) ensuring the nonstationary behavior of dp. [18]

# REFERENCES

---

[17] It is also well known that existing breaks will lower the power of unit root tests (Perron (1989)), thus making stationary processes with breaks difficult to distinguish from those including a unit root.
[18] Standard errors of β are obtained via a non-linear algorithm that involves numerical differentiation.

## Table 1a: Summary Statistics

We present the summary statistics for annual returns, equity premia, risk free rates, dividend-price ratios (dp and $d^*p$) and modified dividend-price ratios (mdp and mdp´). The table shows the correlation matrix among the series as well as the mean, standard deviation and the autocorellation coefficient based on AR(1) fitted model. Data are annual from 1926 to 2012.

|        | $r_t$ | $re_t$ | $rf_t$ | $d^*p_t$ | $dp_t$ | $mdp_t$ | $mdp'_t$ | Mean  | Std  | AR(1) |
|--------|-------|--------|--------|----------|--------|---------|----------|-------|------|-------|
| $r_t$  | 1     |        |        |          |        |         |          | 0.09  | 0.20 | 0.06  |
| $re_t$ | 0.99  | 1      |        |          |        |         |          | 0.06  | 0.20 | 0.05  |
| $rf_t$ | 0.03  | -0.12  | 1      |          |        |         |          | 0.04  | 0.03 | 0.93  |
| $dp_t$ | -0.25 | -0.24  | -0.05  | 0.97     | 1      |         |          | -3.35 | 0.45 | 0.87  |
| $d^*p_t$ | -0.03 | -0.02 | -0.05 | 1       |        |         |          | -3.30 | 0.43 | 0.93  |
| $mdp_t$ | -0.34 | -0.39 | 0.35  | 0.62     | 0.69   | 1       |          | -2.05 | 0.26 | 0.70  |
| $mdp'$ | -0.33 | -0.36  | 0.21   | 0.82     | 0.88   | 0.95    |          | -2.40 | 0.29 | 0.73  |

## Table 1b: Cointegration Test and the null Hypothesis of [1,-1]

We apply the Johansen testing procedure assuming trending series and no trend in the cointegration relationship. The pair [d p] tests for a cointegration relationship between the 12-month summed-up dividends and prices. The 2nd panel presents results for the restriction test that [1 -1] spans the cointegration space between d and p. As usual, (*) and (**) denote rejection at the 5% and 1% levels respectively. Data are overlapping annual spanning the period, 1926-2012.

|         | #coint.vec | Trace Test [d p] | 5% critical value | a) |
|---------|------------|------------------|-------------------|----|
| Panel A | 0          | 19.35*           | 15.49             |    |
|         | <=1        | 0.24             | 3.84              |    |
| Panel B | H0: [1 -1] | $\chi^2$-stat    |                   |    |
|         |            | 10.42**          |                   |    |

# Table 2a: Predictability of returns, equity premia and dividend growth

Standard errors are GMM corrected. Data are annualized constructed from monthly observations with an overlapping rolling window from 1926 to 2012 for Panel A and from **1965 to 2012 for Panel B** respectively.

**Panel A**

|  |  | b | t(b) | $R^2$ |  | b | t(b) | $R^2$ |
|---|---|---|---|---|---|---|---|---|
| $r_t(1)$ | $d^*p_t$ | 0.10 | 2.06 | 0.04 | $r_t(3)$ | 0.27 | 3.43 | 0.11 |
|  | $dp_t$ | 0.09 | 1.57 | 0.03 |  | 0.27 | 3.10 | 0.11 |
|  | $mdp_t$ | 0.21 | 2.84 | 0.07 |  | 0.65 | 5.03 | 0.23 |
|  | $mdp'_t$ | 0.18 | 2.49 | 0.06 |  | 0.56 | 5.63 | 0.21 |
| $r_t(5)$ | $d^*p_t$ | 0.41 | 4.00 | 0.17 | $r_t(7)$ | 0.52 | 3.72 | 0.25 |
|  | $dp_t$ | 0.42 | 4.19 | 0.19 |  | 0.51 | 3.48 | 0.25 |
|  | $mdp_t$ | 1.04 | 9.23 | 0.41 |  | 1.16 | 12.51 | 0.49 |
|  | $mdp'_t$ | 0.89 | 8.08 | 0.36 |  | 1.02 | 8.84 | 0.45 |
| $re_t(1)$ | $d^*p_t$ | 0.10 | 2.11 | 0.04 | $re_t(3)$ | 0.28 | 3.19 | 0.12 |
|  | $dp_t$ | 0.09 | 1.67 | 0.04 |  | 0.28 | 3.19 | 0.12 |
|  | $mdp_t$ | 0.17 | 2.20 | 0.04 |  | 0.53 | 3.17 | 0.16 |
|  | $mdp'_t$ | 0.16 | 2.19 | 0.05 |  | 0.50 | 3.95 | 0.17 |
| $re_t(5)$ | $d^*p_t$ | 0.42 | 3.33 | 0.19 | $re_t(7)$ | 0.56 | 3.15 | 0.29 |
|  | $dp_t$ | 0.44 | 3.67 | 0.21 |  | 0.55 | 3.23 | 0.30 |
|  | $mdp_t$ | 0.83 | 4.19 | 0.26 |  | 0.89 | 4.52 | 0.29 |
|  | $mdp'_t$ | 0.77 | 4.73 | 0.28 |  | 0.87 | 4.94 | 0.34 |
| $\Delta d_t(1)$ | $d^*p_t$ | -0.00 | -0.07 | 0.00 | $\Delta d_t(3)$ | -0.01 | -0.08 | 0.00 |
|  | $dp_t$ | 0.02 | 0.63 | 0.00 |  | 0.01 | 0.13 | 0.00 |
|  | $mdp_t$ | 0.08 | 1.28 | 0.02 |  | 0.08 | 0.71 | 0.01 |
|  | $mdp'_t$ | 0.07 | 1.13 | 0.01 |  | 0.06 | 0.50 | 0.00 |
| $\Delta d_t(5)$ | $d^*p_t$ | -0.01 | -0.11 | 0.00 | $\Delta d_t(7)$ | -0.03 | -0.30 | 0.00 |
|  | $dp_t$ | 0.04 | 0.50 | 0.00 |  | 0.00 | 0.04 | 0.00 |
|  | $mdp_t$ | 0.18 | 1.23 | 0.02 |  | 0.12 | 0.93 | 0.01 |
|  | $mdp'_t$ | 0.13 | 1.16 | 0.01 |  | 0.07 | 0.67 | 0.00 |

**Panel B**

|  |  | b | t(b) | $R^2$ |  | b | t(b) | $R^2$ |
|---|---|---|---|---|---|---|---|---|
| $r_t(1)$ | $d^*p_t$ | 0.11 | 2.08 | 0.09 | $r_t(3)$ | 0.31 | 4.11 | 0.22 |
|  | $dp_t$ | 0.11 | 1.99 | 0.08 |  | 0.32 | 3.67 | 0.23 |
|  | $mdp_t$ | 0.23 | 2.8 | 0.14 |  | 0.62 | 6.68 | 0.36 |
|  | $mdp'_t$ | 0.19 | 2.54 | 0.12 |  | 0.52 | 5.43 | 0.32 |
| $r_t(5)$ | $d^*p_t$ | 0.5 | 7.15 | 0.36 | $r_t(7)$ | 0.67 | 9.63 | 0.47 |
|  | $dp_t$ | 0.51 | 9.55 | 0.37 |  | 0.67 | 8.53 | 0.47 |
|  | $mdp_t$ | 0.97 | 22.99 | 0.54 |  | 1.27 | 12.85 | 0.72 |
|  | $mdp'_t$ | 0.82 | 35.22 | 0.49 |  | 1.07 | 12.63 | 0.65 |
| $re_t(1)$ | $d^*p_t$ | 0.07 | 1.13 | 0.03 | $re_t(3)$ | 0.16 | 1.84 | 0.07 |
|  | $dp_t$ | 0.07 | 1.15 | 0.03 |  | 0.18 | 1.89 | 0.08 |
|  | $mdp_t$ | 0.17 | 1.97 | 0.07 |  | 0.44 | 3.32 | 0.20 |
|  | $mdp'_t$ | 0.13 | 1.69 | 0.06 |  | 0.34 | 2.86 | 0.16 |
| $re_t(5)$ | $d^*p_t$ | 0.26 | 9.34 | 0.12 | $re_t(7)$ | 0.35 | 4.56 | 0.17 |
|  | $dp_t$ | 0.27 | 5.12 | 0.13 |  | 0.35 | 4.27 | 0.17 |
|  | $mdp_t$ | 0.64 | 6.78 | 0.30 |  | 0.86 | 5.04 | 0.44 |
|  | $mdp'_t$ | 0.5 | 9.23 | 0.24 |  | 0.68 | 5.46 | 0.34 |
| $\Delta d_t(1)$ | $d^*p_t$ | 0.04 | 0.86 | 0.01 | $\Delta d_t(3)$ | 0.07 | 0.73 | 0.03 |
|  | $dp_t$ | 0.07 | 1.88 | 0.05 |  | 0.12 | 1.18 | 0.07 |
|  | $mdp_t$ | 0.13 | 2.34 | 0.07 |  | 0.18 | 1.29 | 0.07 |
|  | $mdp'_t$ | 0.12 | 2.18 | 0.07 |  | 0.16 | 1.25 | 0.07 |
| $\Delta d_t(5)$ | $d^*p_t$ | 0.08 | 0.98 | 0.02 | $\Delta d_t(7)$ | 0.07 | 0.85 | 0.02 |
|  | $dp_t$ | 0.13 | 1.57 | 0.07 |  | 0.12 | 1.35 | 0.06 |
|  | $mdp_t$ | 0.18 | 1.57 | 0.05 |  | 0.14 | 1.24 | 0.03 |
|  | $mdp'_t$ | 0.17 | 1.58 | 0.06 |  | 0.14 | 1.31 | 0.04 |

## Table 2b: Univariate forecasting of long run risk free rates

We run univariate regressions among long run risk free rates, $rf_t(h) = \sum_{j=1}^{h} rf_{t+j}$, with the competing dividend-price ratios $(d^*p_t, mdp_t)$ as regressors. Data are annualized constructed from monthly observations with an overlapping rolling window from 1926 to 2012 for Panel (A) and **from 1965 to 2012 for Panel (B).** Standard errors are GMM corrected.

|  | Panel A | | | |  | Panel B | | |
|---|---|---|---|---|---|---|---|---|
|  |  | b | t(b) | $R^2$ |  | b | t(b) | $R^2$ |
| $rf_t(1)$ | $d^*p_t$ | -0.00 | -0.31 | 0.00 | $rf_t(1)$ | $d^*p_t$ 0.05 | 5.55 | 0.48 |
|  | $mdp_t$ | 0.04 | 2.71 | 0.11 |  | $mdp_t$ 0.06 | 4.18 | 0.29 |
| $rf_t(3)$ | $d^*p_t$ | -0.01 | -0.21 | 0.00 | $rf_t(3)$ | $d^*p_t$ 0.15 | 4.99 | 0.60 |
|  | $mdp_t$ | 0.12 | 2.02 | 0.13 |  | $mdp_t$ 0.18 | 3.52 | 0.35 |
| $rf_t(5)$ | $d^*p_t$ | -0.02 | -0.23 | 0.00 | $rf_t(5)$ | $d^*p_t$ 0.24 | 5.70 | 0.73 |
|  | $mdp_t$ | 0.21 | 2.08 | 0.15 |  | $mdp_t$ 0.33 | 4.68 | 0.53 |
| $rf_t(7)$ | $d^*p_t$ | -0.04 | -0.34 | 0.01 | $rf_t(7)$ | $d^*p_t$ 0.32 | 6.29 | 0.72 |
|  | $mdp_t$ | 0.27 | 2.02 | 0.14 |  | $mdp_t$ 0.41 | 4.83 | 0.49 |

## Table 2c: Multivariate predictability of realized returns

This table presents the results of return (and equity premia) predictability for S&P 500 based on the following forecasting regression,

$$r_t(h) = \sum_{j=1}^{h} r_{t+j} = a + b\, dp + c\, mdp + u_t(h)$$

The left-hand variable is the time-$t$ future log return (r) (or equity premium, re) for three, five and seven years ahead ($h = 3,5,7$). The predictor variables include both classical and modified dp ratios (dp and mdp). We use overlapping monthly data in order to formulate the corresponding series for horizons greater than one month, and thus standard errors are GMM corrected. Data are annualized constructed from monthly observations with an overlapping rolling window from 1926 to 2012 for Panel A and from 1965 to 2012 for Panel B.

| Panel A | b(dp) | t(dp) | c(mdp) | t(mdp) | $R^2$ |
|---|---|---|---|---|---|
| $r_t(3)$ | -0.02 | -0.08 | 0.67 | 2.29 | 0.23 |
| $r_t(5)$ | -0.06 | -0.28 | 1.11 | 3.61 | 0.41 |
| $r_t(7)$ | -0.03 | -0.11 | 1.20 | 3.90 | 0.49 |
| $re_t(3)$ | 0.11 | 0.63 | 0.40 | 1.30 | 0.17 |
| $re_t(5)$ | 0.18 | 0.81 | 0.61 | 1.68 | 0.28 |
| $re_t(7)$ | 0.32 | 1.04 | 0.50 | 1.27 | 0.34 |
| Panel B | | | | | |
| $r_t(3)$ | -0.23 | -0.78 | 0.95 | 2.48 | 0.38 |
| $r_t(5)$ | -0.42 | -1.18 | 1.58 | 3.07 | 0.58 |
| $r_t(7)$ | -0.71 | -2.86 | 2.28 | 6.61 | 0.79 |
| $re_t(3)$ | -0.45 | -1.84 | 1.08 | 3.50 | 0.29 |
| $re_t(5)$ | -0.77 | -2.40 | 1.75 | 3.64 | 0.45 |
| $re_t(7)$ | -1.26 | -5.30 | 2.66 | 7.66 | 0.73 |

# Table 3: Univariate analysis based on the CS approximation: Variability of the dp ratio

We run weighted forecasting regressions based on the following univariate type,

$$a + b\, d^*p_t + u_t(h)$$

The left-hand variable represents weighted long-run returns $wr_t(h) = \sum_{j=1}^{h} \rho^{j-1} r_{t+j}$ or long-run dividend growth $wg_t(h) = \sum_{j=1}^{h} \rho^{j-1} \Delta d^*_{t+j}$, or the future dividend-price ratio for five and seven years ahead. To retain consistency dividends here are always reinvested ($d^*$). Data are annual from 1926 to 2012 for Panel A, and from 1965 to 2012 for Panel B. Standard errors are GMM corrected.

Panel A

|  | b | t(b) | $R^2$ |  | b | t(b) | $R^2$ |
|---|---|---|---|---|---|---|---|
| $wr_t(5)$ | 0.37 | 3.84 | 0.17 | $wr_t(7)$ | 0.46 | 3.66 | 0.25 |
| $wg_t(5)$ | -0.01 | -0.12 | 0.00 | $wg_t(7)$ | -0.03 | -0.33 | 0.00 |
| $(0.96^5)\, d^*p_{t+5}$ | 0.60 | 11.30 | 0.50 | $(0.96^7)\, d^*p_{t+7}$ | 0.47 | 9.74 | 0.36 |

Panel B

|  | b | t(b) | $R^2$ |  | b | t(b) | $R^2$ |
|---|---|---|---|---|---|---|---|
| $wr_t(5)$ | 0.46 | 8.67 | 0.34 | $wr_t(7)$ | 0.60 | 11.14 | 0.47 |
| $wg_t(5)$ | 0.07 | 0.77 | 0.03 | $wg_t(7)$ | 0.07 | 1.14 | 0.03 |
| $(0.96^5)\, d^*p_{t+5}$ | 0.57 | 9.47 | 0.47 | $(0.96^7)\, d^*p_{t+7}$ | 0.41 | 7.97 | 0.27 |

## Table 4: Multivariate analysis based on the CS approximation

We run weighted forecasting regressions based on the following multivariate model,

$$a + b \cdot d^*p_t + c \cdot mdp_t + \epsilon_t(h)$$

We also run the same regression with the modified ratio $mdp'_t$ estimated via the ADL methodology. The left-hand variable represents either the weighted realized returns $wr_t(h)$, or weighted dividend growth $wg_t(h) = \sum_{j=1}^{h} \rho^{j-1} \Delta d^*_{t+j}$, or the long run dividend-price ratio for a horizon h of five or seven years ahead. Standard errors are GMM corrected. Data are annual from 1926 to 2012 for Panel A and from 1965 to 2012 for Panel B.

Panel A

|  | slope | t-stat | $R^2$ | slope | t-stat | $R^2$ | slope | t-stat | $R^2$ |
|---|---|---|---|---|---|---|---|---|---|
|  | $wr_t(5)$ | | | $wg_t(5)$ | | | $0.96^5 \cdot d^*p_{t+5}$ | | |
| $d^*p_t$ | 0.01 | 0.05 | 0.37 | -0.11 | -0.74 | 0.03 | 0.85 | 11.69 | 0.62 |
| $mdp_t$ | 0.87 | 3.89 | | 0.24 | 1.02 | | -0.63 | -5.57 | |
| $d^*p_t$ | -0.21 | -0.95 | 0.35 | -0.18 | -0.94 | 0.03 | 1.00 | 9.79 | 0.60 |
| $mdp'_t$ | 1.02 | 3.98 | | 0.31 | 1.14 | | -0.71 | -4.94 | |
|  | $wr_t(7)$ | | | $wg_t(7)$ | | | $0.96^7 \cdot d^*p_{t+7}$ | | |
| $d^*p_t$ | 0.06 | 0.34 | 0.47 | -0.11 | -0.69 | 0.02 | 0.78 | 10.36 | 0.56 |
| $mdp_t$ | 0.93 | 4.37 | | 0.20 | 0.94 | | -0.73 | -6.30 | |
| $d^*p_t$ | -0.15 | -0.64 | 0.43 | -0.17 | -0.87 | 0.02 | 0.93 | 8.55 | 0.51 |
| $mdp'_t$ | 1.05 | 4.20 | | 0.25 | 1.05 | | -0.80 | -5.33 | |

Panel B

|  | slope | t-stat | $R^2$ | slope | t-stat | $R^2$ | slope | t-stat | $R^2$ |
|---|---|---|---|---|---|---|---|---|---|
|  | $wr_t(5)$ | | | $wg_t(5)$ | | | $0.96^5 \cdot d^*p_{t+5}$ | | |
| $d^*p_t$ | -0.36 | -1.33 | 0.57 | -0.16 | -1.09 | 0.08 | 1.15 | 6.87 | 0.56 |
| $mdp_t$ | 1.40 | 4.03 | | 0.39 | 2.40 | | -0.98 | -3.98 | |
| $d^*p_t$ | -0.72 | -2.30 | 0.56 | -0.43 | -2.22 | 0.15 | 1.24 | 5.37 | 0.53 |
| $mdp'_t$ | 1.69 | 4.94 | | 0.71 | 3.59 | | -0.97 | -3.06 | |
|  | $wr_t(7)$ | | | $wg_t(7)$ | | | $0.96^7 \cdot d^*p_{t+7}$ | | |
| $d^*p_t$ | -0.52 | -2.20 | 0.78 | -0.10 | -0.63 | 0.06 | 1.32 | 6.28 | 0.53 |
| $mdp_t$ | 1.85 | 6.09 | | 0.28 | 1.20 | | -1.50 | -4.83 | |
| $d^*p_t$ | -0.91 | -2.47 | 0.73 | -0.32 | -1.50 | 0.10 | 1.50 | 4.79 | 0.44 |
| $mdp'_t$ | 2.13 | 4.83 | | 0.54 | 2.02 | | -1.53 | -3.67 | |

## Table 4b: Multivariate slope breakdown analysis based on CS approximation

We run weighted forecasting regressions based on the following multivariate model,

$$a + b \cdot d^*p_t + c \cdot mdp_t + \epsilon_t(h)$$

We also run the same regression with the modified ratio $mdp'_t$ estimated via the ADL methodology. The left-hand variable represents either the weighted realized equity premia $wre_t(h) = \sum_{j=1}^{h} \rho^{j-1} re_{t+j}$ or risk free returns $wrf_t(h) = \sum_{j=1}^{h} \rho^{j-1} rf_{t+j}$ for a horizon h of five or seven years ahead. Standard errors are GMM corrected. Data are annual from 1926 to 2012 for Panel A and from 1965 to 2012 for Panel B.

### Panel A

|         | $wre_t(5)$ | $wre_t(7)$ |
|---------|-----------|-----------|
| $d^*p_t$ | 0.17 | 0.31 |
| $mdp_t$  | 0.51 | 0.43 |
| $d^*p_t$ | 0.03 | 0.21 |
| $mdp'_t$ | 0.62 | 0.50 |

|         | $wrf_t(5)$ | $wrf_t(7)$ |
|---------|-----------|-----------|
| $d^*p_t$ | -0.17 | -0.25 |
| $mdp_t$  | 0.37 | 0.50 |
| $d^*p_t$ | -0.25 | -0.36 |
| $mdp'_t$ | 0.40 | 0.55 |

### Panel B

|         | $wre_t(5)$ | $wre_t(7)$ |
|---------|-----------|-----------|
| $d^*p_t$ | -0.69 | -0.96 |
| $mdp_t$  | 1.56 | 2.09 |
| $d^*p_t$ | -1.08 | -1.37 |
| $mdp'_t$ | 1.88 | 2.37 |

|         | $wrf_t(5)$ | $wrf_t(7)$ |
|---------|-----------|-----------|
| $d^*p_t$ | 0.33 | 0.43 |
| $mdp_t$  | -0.16 | -0.24 |
| $d^*p_t$ | 0.36 | 0.46 |
| $mdp'_t$ | -0.19 | -0.24 |

**Table 5: Out of Sample (OS) evaluation**

We present OS results for classical and the two modified dp ratios: one estimated with a recursive procedure mdp(R) and one where the entire sample is used to estimate the cointegrating coefficient mdp(P). Data are overlapping annual spanning the period 1926 to 2012 for Panel A and 1965 to 2012 for Panel B.

.

**Panel A**

| Realized returns | r(3) | r(5) | r(7) |
|---:|---|---|---|
| dp | -0.03 | -0.02 | 0.14 |
| d*p | -0.15 | -0.05 | 0.13 |
| mdp (P) | 0.34 | 0.49 | 0.49 |
| mdp'(P) | 0.27 | 0.38 | 0.43 |
| mdp (R) | 0.07 | 0.26 | 0.31 |

| Realized premia | re (3) | re (5) | re (7) |
|---:|---|---|---|
| dp | -0.01 | 0.00 | 0.16 |
| d*p | -0.12 | -0.02 | 0.15 |
| mdp (P) | 0.20 | 0.28 | 0.26 |
| mdp'(P) | 0.20 | 0.29 | 0.32 |
| mdp (R) | -0.11 | -0.01 | 0.03 |

**Panel B**

| Realized returns | r(3) | r(5) | r(7) |
|---:|---|---|---|
| dp | -0.20 | 0.01 | -0.10 |
| $d^*p$ | -0.11 | 0.08 | -0.07 |
| mdp (P) | 0.26 | 0.52 | 0.66 |
| mdp´(P) | 0.15 | 0.40 | 0.47 |
| mdp (R) | -0.14 | 0.07 | 0.28 |

| Realized premia | re(3) | re(5) | re(7) |
|---:|---|---|---|
| dp | -0.39 | -0.24 | -0.41 |
| $d^*p$ | -0.33 | -0.19 | -0.35 |
| mdp (P) | 0.07 | 0.25 | 0.28 |
| mdp´(P) | -0.05 | 0.13 | 0.05 |
| mdp (R) | -0.18 | -0.10 | -0.02 |

**Figure 1: Evolution of dp and mdp against forward looking 5-year returns.**

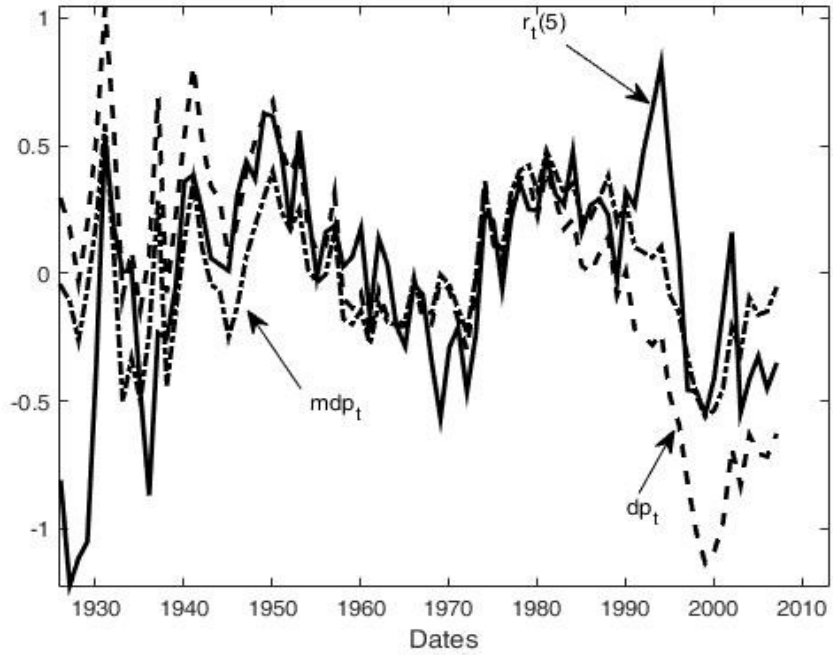

**Figure 2: Convergence of a recursively estimated cointegration b coefficient to its population value β**

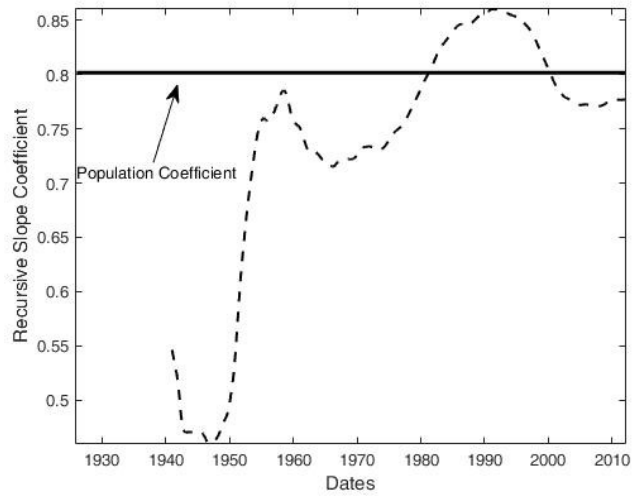